\documentclass[aps,prb,showpacs,floatfix,twocolumn,amsmath,amssymb,preprintnumbers]{revtex4-1}
\usepackage{mathrsfs}
\usepackage[figuresright]{rotating}
\usepackage{amsmath}
\usepackage{amssymb}
\usepackage{graphicx}
\usepackage{color}
\usepackage{dcolumn}
\usepackage{bm}
\usepackage[breaklinks=true,colorlinks=true,linkcolor=blue,urlcolor=blue,citecolor=blue]{hyperref}

\makeatletter
\newcommand{\rmnum}[1]{\romannumeral #1}
\newcommand{\Rmnum}[1]{\expandafter\@slowromancap\romannumeral #1@}
\makeatother

\begin{document}

\title{Electron-hole compensation effect between topologically trivial electrons and nontrivial holes in NbAs}

\author{Yongkang Luo}
\email[]{ykluo@lanl.gov}
\author{N. J. Ghimire}
\author{M. Wartenbe}
\author{Hongchul Choi}
\author{M. Neupane}
\author{R. D. McDonald}
\author{E. D. Bauer}
\author{Jianxin Zhu}
\author{J. D. Thompson}
\author{F. Ronning}
\email[]{fronning@lanl.gov}
\affiliation{Los Alamos National Laboratory, Los Alamos, New Mexico 87545, USA.}


\date{\today}

\begin{abstract}

Via angular Shubnikov-de Hass (SdH) quantum oscillations measurements, we determine the Fermi surface topology of NbAs, a Weyl semimetal candidate. The SdH oscillations consist of two frequencies, corresponding to two Fermi surface extrema: 20.8 T ($\alpha$-pocket) and 15.6 T ($\beta$-pocket). The analysis, including a Landau fan plot, shows that the $\beta$-pocket has a Berry phase of $\pi$ and a small effective mass $\sim$0.033 $m_0$, indicative of a nontrivial topology in momentum space; whereas the $\alpha$-pocket has a trivial Berry phase of 0 and a heavier effective mass $\sim$0.066 $m_0$. From the effective mass and the $\beta$-pocket frequency we determine that the Weyl node is 110.5 meV from the chemical potential. A novel electron-hole compensation effect is discussed in this system, and its impact on magneto-transport properties is addressed. The difference between NbAs and other monopnictide Weyl semimetals is also discussed.

\end{abstract}


\maketitle

\section{Introduction}

It has long been realized that the band structure of a material could conspire such that energy bands intersect at individual points in momentum space with a dispersion that can be described by massless Dirac particles \cite{Herring1937}. If the material possesses time reversal symmetry, inversion symmetry, and a small carrier density due to the chemical potential lying close to the Dirac point, the material is known as a Dirac semimetal. Recent predictions of such materials have been confirmed experimentally by photoemission \cite{Young-3DDSM,Wang-Cd3As2DSM,Wang-A3BiDSM,Neupane-Cd3As2ARPES,Liu-Na3Bi}. However, if either symmetry is broken, the doubly degenerate Dirac point can split into a pair of Weyl nodes with opposite chirality, and the material is referred to as a Weyl semimetal. The topological nature of these materials was recently recognized, with pairs of Weyl nodes acting as sources and drains of Chern flux \cite{Balents-Weyl,Hosur-WSM,Wan-5dSpinel,Wang-WSM2013}. As in the case of topological insulators, the topological aspects of Weyl semimetals can manifest themselves with novel states (Fermi arcs in this case) at the surface of the material\cite{Hosur-Friedel,Ojanen-WSMFmArc,Potter-QOFmArc}. Other exotic phenomena associated with Weyl semimetals include a chiral anomaly, non-local transport, and the quantum anomalous Hall effect \cite{Zyuzin-TopRes,Wang-WSM2013,Parameswaran-Nonlocal,Burkov-JPCM2015}.

Early proposals of Weyl semimetals exploited broken time reversal symmetry \cite{Xu-HgCr2Se4,Burkov-WSM2011,Wan-5dSpinel}, but recently a class of binary transition-metal monopnictides of the form $TmPn$ (where $Tm$ = Ta or Nb, and $Pn$ = As or P) have been predicted to be Weyl semimetals on the basis of their broken spatial inversion symmetry \cite{Weng-TmPn,Huang-TaAsInv}. These materials possess extremely high mobilities \cite{Zhang-TaAsSdH,Huang-TaAsLMR,Shekhar-NbP,Nirmal-NbAs}, and evidence for their topological nature has been presented by angle-resolved photoemission spectroscopy (ARPES) experiments \cite{XuS-TaAsARPES,Lv-TaAsPRX,Lv-TaAsNP}. Due to the very high mobilities, quantum oscillations can be readily observed, which enables a detailed examination of the bulk 3D electronic structure of the material.

In this work we provide a detailed study of the electronic structure of NbAs through Shubnikov-de Hass (SdH) oscillations observed in the transverse magnetoresistance. The small carrier density with both electrons and holes is evident from the small frequencies and light effective masses. In addition, from an analysis including a Landau fan and Hall effect, we conclude that the pocket we assigned as hole-like is topological in nature, while the other pocket assigned to be electron-like is not. The large ratio between transport life time and quantum life time (hole pocket), $\tau_{tr}/\tau_{Q}\sim$ 1000, suggests a remarkable topological protection mechanism that strongly suppresses backward scattering in zero magnetic field. These results demonstrate a novel compensation effect between topologically trivial electrons and nontrivial holes in NbAs, and the exotic transport properties of this material are consequences of a dual effect of electron-hole compensation and topological protection, which is distinguished from the parabolic semimetals (e.g., Bi\cite{Alers-BiMR} and WTe$_2$\cite{Ali-WTe2XMR}), the well-studied Dirac semimetals Cd$_3$As$_2$\cite{Ong-Cd3As2}, and even its analogous Weyl semimetal TaAs\cite{Zhang-TaAsSdH}.

\section{Experimental details}

Millimeter sized single crystals of NbAs were synthesized by a vapor transport method with iodine as described elsewhere\cite{Nirmal-NbAs}. A high quality crystal (residual resistance ratio [$RRR\equiv R(300 K)/R(2 K)$] $=$ 72) was oriented by checking high symmetry reflections using X-ray diffraction (XRD). Ohmic contacts were prepared on the plate-like NbAs crystal in a Hall-bar geometry, and both in-plane electrical resistivity ($\rho_{xx}$) and Hall resistivity ($\rho_{yx}$) were measured by slowly sweeping a DC magnetic field from $-$18 T to 18 T at a rate of 0.2 T/min. $\rho_{xx}$ ($\rho_{yx}$) was obtained as the symmetric (antisymmetric) component under magnetic field reversal. An AC-resistance bridge (LR-700) was used to perform these transport measurements in a He-3 refrigerator.

\section{Results and Discussion}

\subsection{SdH oscillations for $\textbf{B}\parallel\textbf{c}$}

\begin{figure}[htbp]
\vspace*{-25pt}
\includegraphics[width=9cm]{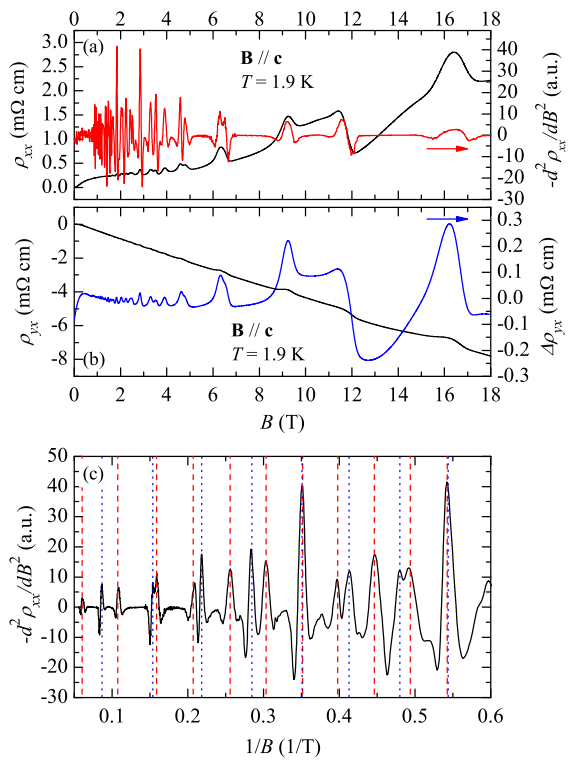}
\vspace*{-25pt}
\caption{\label{Fig.1} Magnetic field dependence of (a) $\rho_{xx}$ and (b) $\rho_{yx}$ at 1.9 K. The right axes respectively show $-d^2\rho_{xx}/dB^2$ and $\Delta \rho_{yx} = \rho_{yx}-\langle\rho_{yx}\rangle$. Panel (c) displays $-d^2\rho_{xx}/dB^2$ as a function of $1/B$. }
\end{figure}

NbAs is highly metallic in the absence of magnetic field \cite{Nirmal-NbAs}. When a magnetic field is present, $\textbf{B}\parallel\textbf{c}$, the resistivity increases rapidly and shows an ultrahigh magnetoresistance [\%$ MR \equiv 100\times(\rho_{xx}(B)-\rho_{xx}(0))/\rho_{xx}(0)$], [Fig.~\ref{Fig.1}(a)]. At 1.9 K and 18 T, the $MR$ reaches 462,000\%, while little evidence of saturation can be seen. This value of $MR$ is comparable with other Weyl semimetal candidates TaAs\cite{Zhang-TaAsSdH,XuS-TaAsARPES,Huang-TaAsLMR} and NbP\cite{Shekhar-NbP}, and is characteristic of a high mobility. The Hall effect changes sign as a function of temperature from hole-like at high temperatures to electron-like at low temperatures \cite{Nirmal-NbAs}. In Fig.~\ref{Fig.1}(b), we present our Hall resistivity $\rho_{yx}$ as a function of $B$ at 1.9 K. At high magnetic field, $\rho_{yx}$ is almost a linear function of $B$ superposed with a large SdH-oscillation signal. Such a $\rho_{yx}(B)$ profile manifests the coexistence of hole carriers with smaller concentration and electron carriers with larger concentration (See below). We will see that this is consistent with the SdH-oscillation analysis discussed hereafter.

\begin{figure}[htbp]
\vspace*{-20pt}
\includegraphics[width=8.5cm]{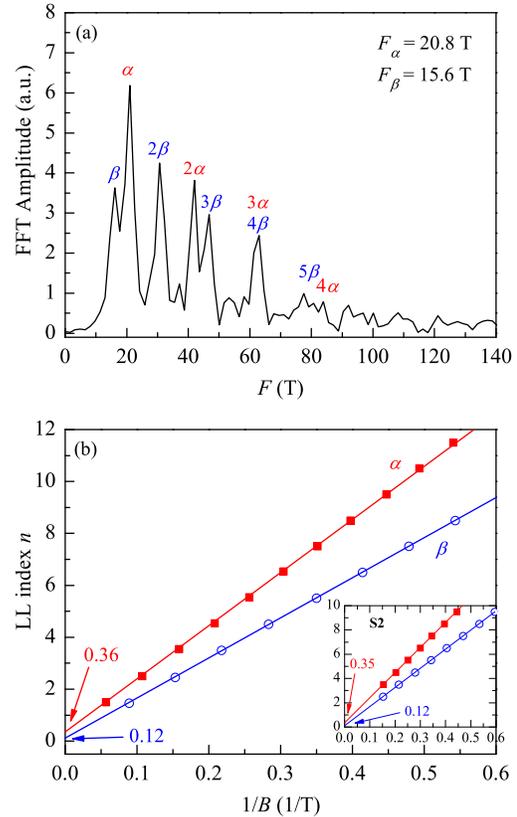}
\vspace*{-10pt}
\caption{\label{Fig.2} (a) FFT spectrum of $-d^2\rho_{xx}/dB^2$, showing two pronounced frequencies, $F_{\alpha}=$20.8 T, and $F_{\beta}=$15.6 T.
(b) Landau fan diagram. The peaks of $-d^2\rho_{xx}/dB^2$ are assigned as half-integer Landau level indices. The linear extrapolation to the infinite field limit results in the intercepts of $\sim$0.36 and $\sim$0.12 for $\alpha$- and $\beta$-pockets, respectively. The inset to (b) shows the results measured on another sample \textbf{S2}.}
\end{figure}

Another important feature of magneto-transport properties in NbAs is the large SdH quantum oscillations, observed in both $\rho_{xx}(B)$ [Fig.~\ref{Fig.1}(a)] and $\rho_{yx}(B)$ [Fig.~\ref{Fig.1}(b)]. To better resolve the SdH oscillations, we show $-d^2\rho_{xx}/dB^2$ and $\Delta\rho_{yx}=\rho_{yx}-\langle\rho_{yx}\rangle$ in the right axes. Here $\langle\rho_{yx}\rangle$ is the non-oscillatory background of $\rho_{yx}$. The result of $-d^2\rho_{xx}/dB^2$ is also displayed in Fig.~\ref{Fig.1}(c) as a function of $1/B$. It is clearly seen that all the peaks in $-d^2\rho_{xx}/dB^2$ are well indexed by two frequencies as marked with the red-dashed and blue-dotted lines.

\begin{figure}[htbp]
\vspace*{-20pt}
\includegraphics[width=9cm]{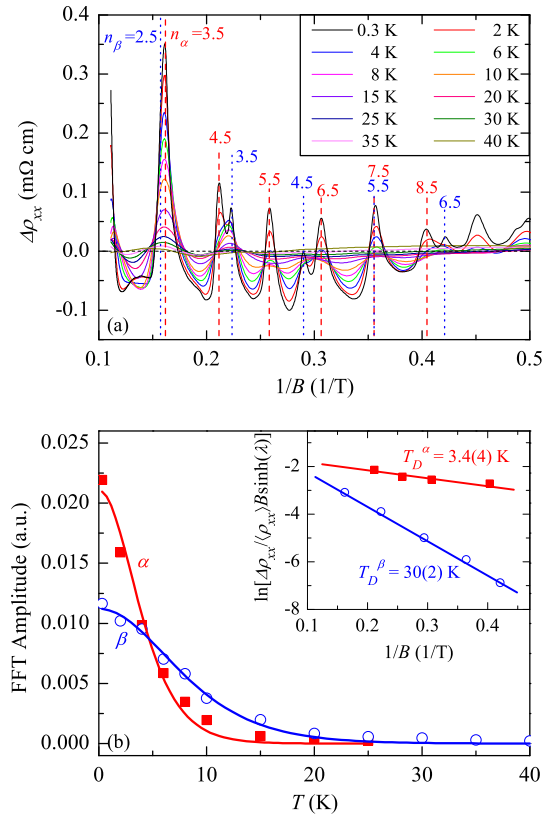}
\vspace*{-20pt}
\caption{\label{Fig.3} (a) SdH oscillations at various temperatures. Here $\Delta \rho_{xx}=\rho_{xx}-\langle\rho_{xx}\rangle$. The red dashed and blue dotted lines mark the peak positions for $\alpha$- and $\beta$-pockets, respectively. (b) The FFT amplitudes of $\alpha$- and $\beta$-orbits as a function of $T$. The solid lines are the fits to Lifshitz-Kosevich formula. The inset is a Dingle plot for the $\beta$-pocket at 15 K, where $\lambda=2\pi^2k_BT/\hbar\omega_c$.}
\end{figure}

In order to analyze the SdH oscillations of $\rho_{xx}$ in NbAs, we use the following expression for a 3D system\cite{Shoenberg,Murakawa-Rashba,Zhang-TaAsSdH,ChenXH-P}:
\begin{equation}
\frac{\Delta\rho_{xx}}{\langle\rho_{xx}\rangle}=A(T,B)\cos[2\pi(\frac{F}{B}-\gamma+\delta)],
\label{Eq.1}
\end{equation}
in which $\langle\rho_{xx}\rangle$ is the non-oscillatory part of $\rho_{xx}$, $F$ is the frequency of oscillation, $\gamma$ is the Onsager phase, and $\delta$ is an additional phase shift taking a value between $\pm1/8$ depending on the curvature of the FS topology\cite{Murakawa-Rashba,WangJ-Cd3As2SdH}.
By performing a fast Fourier transformation (FFT) on $-d^2\rho_{xx}/dB^2$, we derived two oscillation frequencies, $F_{\alpha} =$ 20.8 T, and $F_{\beta} =$ 15.6 T, and their higher-order harmonics, as shown in Fig.~\ref{Fig.2}(a). According to the Lifshitz-Onsager relation, the quantum oscillation frequency $F$ is proportional to the extremal cross-sectional area $S_F$ of the Fermi surface (FS), {\it i.e.},
\begin{equation}
F=\frac{\hbar}{2\pi e}S_F.
\label{Eq.2}
\end{equation}
The magnitudes of Fermi wave vector for $\alpha$- and $\beta$-pockets are then calculated via $k_F=(S_F/\pi)^{1/2}$. The calculated values are $k_F^{\alpha}=0.025$ \AA$^{-1}$ and $k_F^{\beta}=0.022$ \AA$^{-1}$, respectively. We should point out that these cross-sectional areas are very small, only taking up $\sim$0.05\% of the whole area of Brillouin zone in the $\mathbf{k_x}$-$\mathbf{k_y}$ plane. Such small Fermi pockets are responsible for the low carrier concentration (see below), and are consistent with the semimetallic nature of NbAs.

\begin{figure*}[htbp]
\vspace*{-15pt}
\includegraphics[width=2.0\columnwidth]{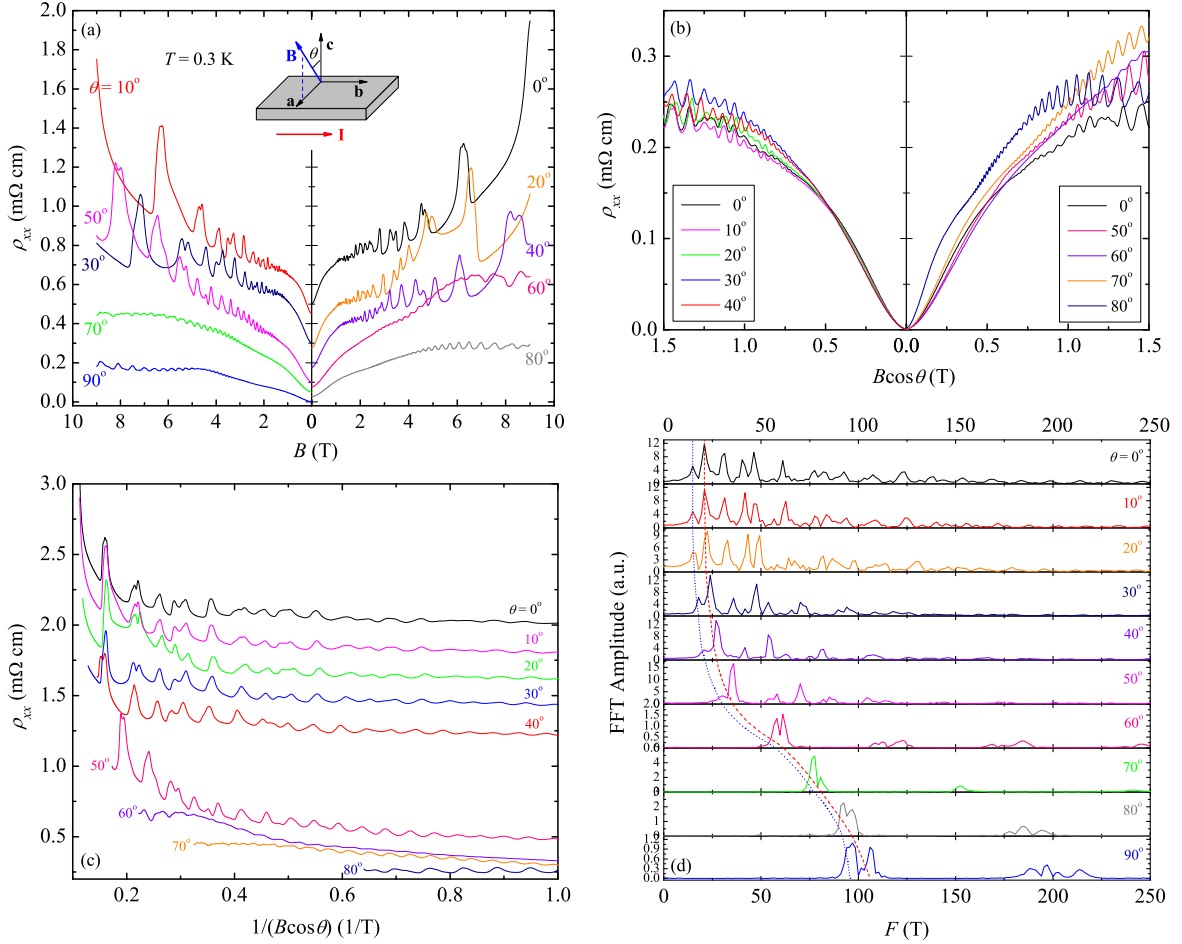}
\vspace*{-10pt}
\caption{\label{Fig.4} (a) Field dependence of $\rho_{xx}$ at selected angles. The data are vertically shifted for clarity. The inset is a schematic sketch of the measurement configuration. (b) Low field $\rho_{xx}$ plotted in the scale of $B\cos\theta$. (c) SdH patterns at various angles plotted as a function of $1/(B\cos\theta)$. (d) Fermi surface extrema evolution with $\theta$. The red and blue lines are guides to eyes for $\alpha$- and $\beta$-pockets, respectively. }
\end{figure*}

The Onsager phase $\gamma=1/2-\Phi_B/2\pi$ with $\Phi_B$ being a Berry phase. In a topologically trivial band with a parabolic dispersion, $\Phi_B$ is zero and therefore $\gamma=1/2$; whereas, in a Dirac electronic system, $\gamma=0$ due to a topologically non-trivial Berry phase $\Phi_B=\pi$. To identify the topological feature of these two pockets, we plot the Landau fan diagram in Fig.~\ref{Fig.2}(b). Since $\langle\rho_{yx}\rangle/\langle\rho_{xx}\rangle\sim7$, $\sigma_{xx}$ is in phase with $\rho_{xx}$\cite{Xiang-BiTeCl}; therefore, the minima and maxima of the SdH oscillations in $\rho_{xx}$ (or $-d^2\rho_{xx}/dB^2$) are assigned as integer ($n$) and half-integer ($n+$1/2) Landau level (LL) indices, respectively.
A linear extrapolation of $n$ versus $1/B$ to the infinite field limit gives rise to the intercepts ($\gamma-\delta$) for $\alpha$- and $\beta$-pockets as shown in Fig.~\ref{Fig.2}(b). We find that the intercept of the $\beta$-pocket is $\sim$0.12, falling between $-$1/8 and 1/8, indicative of a non-trivial $\pi$ Berry phase; while the intercept for the $\alpha$-pocket is $\sim$0.36, close to 0.5 when taking into account the additional phase shift $\delta$, implying a topologically trivial Berry phase 0. Similar results were reproduced on another sample \textbf{S2} as depicted in the inset to Fig.~\ref{Fig.2}(b). This $n$ versus $1/B$ plot also reveals that the system enters the quantum limit when the magnetic field is above 28.4 T (17.2 T) for the $\alpha$-pocket ($\beta$-pocket). These relatively low field values for the quantum-limit field are in agreement with the low carrier density and semimetallicity of NbAs. Furthermore, magnetization measurements made through the quantum limit confirm the topological nature of the $\beta$-pocket\cite{Philip-NbAsMag}.

We turn now to the temperature dependent SdH oscillations shown in Fig.~\ref{Fig.3}(a).
By looking at the peaks $n_{\alpha}=4.5$ and $n_{\beta}=3.5$ [Fig.~\ref{Fig.3}(a)], one clearly finds that the SdH oscillations of the $\alpha$-pocket decay much faster with increasing $T$ than those of the $\beta$-pocket. A similar situation is also seen in the $n_{\alpha}=8.5$ and $n_{\beta}=6.5$ peaks; and this is also confirmed by the peak-valley evolution as $T$ increases at the $n_{\alpha}=5.5$ peak, which also coincides with the valley position of $n_{\beta}=4$. All these qualitatively suggest that the effective mass of the $\alpha$-pocket is heavier than that of the $\beta$-pocket. We also notice that the SdH oscillations of the $\alpha$-pocket nearly disappear when $T$ exceeds 15 K. The decaying amplitude of SdH oscillations with temperature is described by the Lifshitz-Kosevich (LK) formula\cite{Murakawa-Rashba,Zhang-TaAsSdH}:
\begin{equation}
A(T,B)\propto\exp(-2\pi^2k_BT_D/\hbar\omega_c)\frac{2\pi^2k_BT/\hbar\omega_c}{\sinh(2\pi^2k_BT/\hbar\omega_c)},
\label{Eq.3}
\end{equation}
in which $T_D$ is the Dingle temperature, and $\omega_c=\frac{eB}{m^*}$ is the cyclotron frequency with $m^*$ being the effective mass. We tracked the FFT amplitudes of $\alpha$- and $\beta$-orbits as a function of $T$ in Fig.~\ref{Fig.3}(b). Fitting these data points to the LK formula, we derived the effective masses, $m^*_{\alpha}=0.066(5)~m_0$ and $m^*_{\beta}=0.033(2)~m_0$, where $m_0$ is the mass of a free electron. Note that $m^*_{\alpha}$ is twice of $m^*_{\beta}$. The validity of these values is verified by tracking the temperature-dependent FFT amplitudes of the $2\alpha$- and $2\beta$-orbits (data not shown), from which we obtained the effective masses that are nearly doubled, $m^*_{2\alpha}=0.130(9)~m_0$ and $m^*_{2\beta}=0.062(7)~m_0$. We should emphasize that the small value of $m^*_{\beta}$ is comparable with or even smaller than most of the known topological materials, e.g., 0.089 $m_0$ for the 3D topological insulator Bi$_2$Te$_2$Se\cite{Ren-Bi2Te2Se,Ong-Bi2Te2Se}, 0.043 $m_0$ for the 3D Dirac semimetal Cd$_3$As$_2$\cite{WangJ-Cd3As2SdH}, 0.11 $m_0$ for the 3D Dirac semimetal Na$_3$Bi\cite{Ong-Na3BiSdH}, and 0.15 $m_0$ for the 3D Weyl semimetal TaAs\cite{Zhang-TaAsSdH}, which adds to our confidence that the low-energy electronic excitations are due to the massless Weyl fermions in the $\beta$-pocket. A Dingle fit is presented in the inset to Fig.~\ref{Fig.3}(b). The fitted Dingle temperatures for $\alpha$- and $\beta$-pockets are $T_D^{\alpha}=3.4$ K and $T_D^{\beta}=30$ K. From these, we calculate the quantum lifetime $\tau_Q^{\alpha}=\frac{\hbar}{2\pi k_BT_D^{\alpha}}=3.6\times10^{-13}$ s, while the value for $\beta$-pocket is 10 times smaller, $\tau_Q^{\beta}=3.8\times10^{-14}$ s.

From the effective mass, the measured Fermi surface area, and assuming a parabolic (linear) dispersion for the $\alpha$- ($\beta$-) pocket we can determine the Fermi energy as reported in Table \ref{Tab.1}. It is important to note that while the Fermi energy of -110.5 meV for the $\beta$- pocket is small relative to the overall bandwidth of the Nb $4d$-states, it is still large relative to thermal energies up to room temperature. We note that the magnitude of the positive conductivity contribution to the magnetoresistance is inversely proportional to the Fermi energy\cite{Son-ChirAnom}. Hence, the relatively large Fermi energy for our samples may limit the ability to resolve the chiral anomaly, which we could not resolve when performing longitudinal magnetoresistance measurements. One possible reason for this relatively large Fermi energy is due to an off stoichiometric in the sample composition\cite{Nirmal-NbAs}. We should also note that a negative longitudinal magnetoresistance could also be overwhelmed by several competing effects like weak anti-localization or Fermi surface anisotropy\cite{Shekhar-TaPLMR}.

\subsection{Angular dependent SdH oscillations}

Fig.~\ref{Fig.4}(a) shows the field dependence of $\rho_{xx}$ measured at various angles. The configuration of magnetic field $\textbf{B}$ and electrical current $\textbf{I}$ is shown in the inset to Fig.~\ref{Fig.4}(a). For clarity, we have vertically shifted the curves. In Fig.~\ref{Fig.4}(b), we present the low-field region of $\rho_{xx}$ as a function of $B\cos\theta$. When $\theta<45$\textordmasculine, all the curves almost collapse onto a single line for $B<0.5$ T. This scaling is violated when $\theta>45$\textordmasculine. Similar phenomenon is also seen in the high field quantum oscillation regime, for which we have plotted $\rho_{xx}$ versus $1/(B\cos\theta)$ in Fig.~\ref{Fig.4}(c). One clearly sees that the SdH pattern remains essentially unchanged when $\theta<45$\textordmasculine, but gradually deviates as $\theta>45$\textordmasculine. Generically, as $\theta$ increases, the amplitude of SdH oscillations shrinks while the oscillatory frequency increases. These results suggest that the FS topology could be banana-like: they exhibit some 2D-like features at low $\theta$ albeit an overall 3D-like property. To figure out the angular dependent FS topology, we display the FFT spectra in Fig.~\ref{Fig.4}(d) for all $\theta$ between 0\textordmasculine and 90\textordmasculine in steps of 10\textordmasculine. The red-dashed and blue-dotted lines signify the angular evolution of the $\alpha$- and $\beta$-pockets, respectively. Note that for all angles, $S_F^{\alpha}$ is larger than $S_F^{\beta}$, demonstrating that the $\alpha$-pocket is larger than the $\beta$-pocket in $\textbf{k}$-space volume. This manifests that the carriers in the $\alpha$-pocket should be the majority carriers.

\begin{figure}[htbp]
\vspace*{-20pt}
\includegraphics[width=8.5cm]{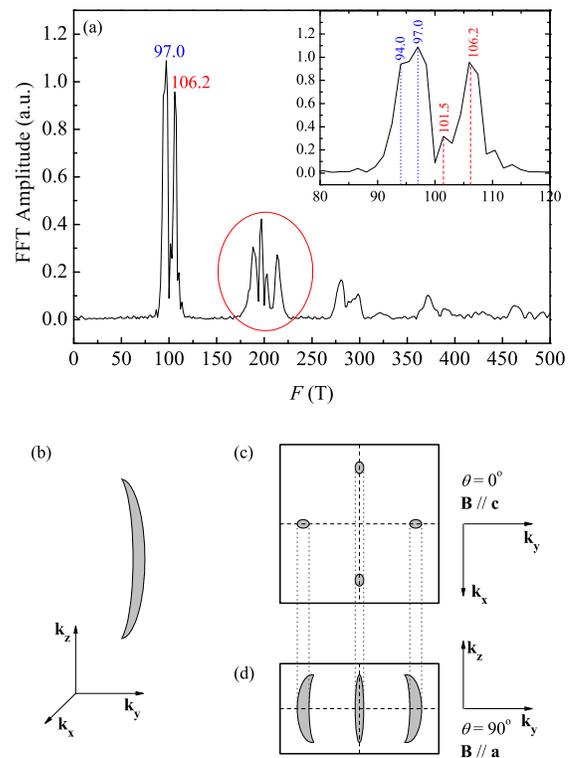}
\caption{\label{Fig.5} SdH oscillations for $\theta=90$\textordmasculine. (a) FFT spectrum of $-d^2\rho_{xx}/dB^2$; the inset is an enlarged plot that shows the four observed frequencies. (b) shows a schematic diagram of a banana-like FS topology. (c) $\theta=0$\textordmasculine, $\textbf{B}$$\parallel$$\textbf{c}$, the banana-like FS only has one cross-sectional extremum. (d) $\theta=90$\textordmasculine, $\textbf{B}$$\parallel$$\textbf{a}$, the banana-like FS has two cross-sectional extrema, corresponding to two different SdH frequencies. }
\end{figure}

We systematically analyzed the SdH-oscillation data for $\theta=90$\textordmasculine. Compared with the configuration of $\theta=0$\textordmasculine, three prominent features are apparent: (\rmnum{1}), the oscillatory amplitude is much smaller than that of $\theta=0$\textordmasculine, whereas (\rmnum{2}), the oscillatory frequency is much higher. This implies larger effective masses but lower carrier mobilities when compared to the $\textbf{B}\parallel\textbf{c}$ configuration, which is consistent with the fact that the $MR$ for $\textbf{B}\parallel\textbf{a}$ is also much smaller [cf. Fig.~\ref{Fig.4}(a)]. (\rmnum{3}), We can also see a very obvious beat pattern in this configuration, implying the existence of multiple frequencies that are very close to each other. Indeed, after performing a FFT on $-d^2\rho_{xx}/dB^2$, we obtained four frequencies, 94.0 T, 97.0 T, 101.5 T, and 106.2 T as shown in the inset to Fig.~\ref{Fig.5}(a). Actually, these four frequencies are much better distinguished in the second harmonic highlighted in the mainframe of Fig.~\ref{Fig.5}(a).

The existence of four frequencies in the configuration $\theta=90$\textordmasculine ($\textbf{B}$$\parallel$$\textbf{a}$) is not surprising, considering an upright banana-like FS topology as shown in Fig.~\ref{Fig.5}(b). When $\theta=0$\textordmasculine ($\textbf{B}$$\parallel$$\textbf{c}$) both $\alpha$- and $\beta$-pockets have only one cross-sectional area $S_F$ in the $\mathbf{k_x}$-$\mathbf{k_y}$ plane [Fig.~\ref{Fig.5}(c)], and this gives rise to the two oscillation frequencies as discussed above. While for $\theta=90$\textordmasculine, both $\alpha$- and $\beta$-pockets have two cross-sectional areas when projected onto the $\mathbf{k_y}$-$\mathbf{k_z}$ plane [Fig.~\ref{Fig.5}(d)]; therefore, it is reasonable that the FFT peaks split in this configuration.

\subsection{Discussions}

\begin{figure}[htbp]
\hspace*{-12pt}
\includegraphics[width=9.5cm]{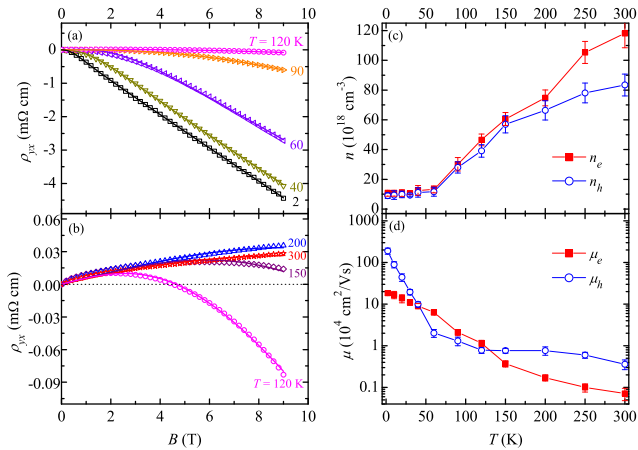}
\caption{\label{Fig.6} Hall effect of NbAs. (a-b) show field dependent $\rho_{yx}$ at selected temperatures. The open symbols are experimental data, and the solid line through the data are numerical fit to a two-band model. The curve for 2 K is the non-oscillating background $\langle\rho_{yx}\rangle$. (c) and (d) display temperature dependencies of carrier density and mobility, respectively.}
\end{figure}

To estimate the carrier density and mobility, we systematically measured the Hall effect at different temperatures. Some selected curves are shown in Figs.~\ref{Fig.6}(a-b). At high temperatures $T>$ 150 K, $\rho_{yx}$ is positive, but at low temperatures the Hall resistivity changes sign and grows significantly\cite{Nirmal-NbAs}. This suggests a multi-band effect. For simplicity, we adopted one electron-band and one hole-band, and we will see that this is sufficient to describe the Hall effect for the full temperature range. By carefully fitting $\rho_{yx}(B)$ to this two-band model\cite{LuoY-WTe2Hall}:
\begin{equation}
\rho_{yx}(B)=\frac{B}{e}\frac{(n_h\mu_h^2-n_e\mu_e^2)+(n_h-n_e)\mu_e^2\mu_h^2B^2}{(n_h\mu_h+n_e\mu_e)^2+[(n_h-n_e)\mu_e\mu_hB]^2},
\label{Eq.4}
\end{equation}
where $n$ and $\mu$ are respectively carrier density and mobility, and the subscript $e$ (or $h$) denotes electron (or hole), we derived the temperature dependencies of carrier density and mobility for both electron- and hole-type carriers, and the results are displayed in Figs.~\ref{Fig.6}(c) and (d), respectively. $\mu_e$ increases continuously from 7.15$\times 10^2$ cm$^2$/Vs at 300 K to 1.9$\times 10^5$ cm$^2$/Vs at 2 K. On the contrary, $\mu_h$ remains essentially unchanged for temperatures above 60 K, but rapidly increases below 60 K, and reaches 1.9$\times10^6$ cm$^2$/Vs at 2 K. Both $n_e$ and $n_h$ decrease monotonically with decreasing temperature, and finally saturate at low temperature. In particular, at 2 K, we find $n_e=1.08\times 10^{19}$cm$^{-3}$ and $n_h=9.04\times 10^{18}$cm$^{-3}$. Such low carrier densities are consistent with the semi-metallicity of NbAs. We should note that the ratio of $n_e/n_h$ is close to the ratio of Fermi volume between $\alpha$- and $\beta$-pockets, $V_{k\alpha}/V_{k\beta}\sim$ 1.26, if we assume that both of $\alpha$- and $\beta$-pockets are elliptical. More physical parameters are calculated and summarized in Table \ref{Tab.1}.

\begin{table}
\tabcolsep 0pt \caption{\label{Tab.1} Physical parameters of $\alpha$- and $\beta$-pockets for $\textbf{B}\parallel\textbf{c}$.}
\vspace*{-12pt}
\begin{center}
\def\temptablewidth{1.0\columnwidth}
{\rule{\temptablewidth}{1pt}}
\begin{tabular*}{\temptablewidth}{@{\extracolsep{\fill}}ccc}
Quantities                         &     $\alpha$                     &     $\beta$                      \\ \hline
$F$ (T)                            &     20.8(2)                      &     15.6(2)                      \\
$k_F$ (\AA$^{-1}$)                 &     0.025(1)                     &     0.022(1)                     \\
Intercept                          &     0.36(1)                      &     0.12(1)                      \\
$\Phi_B$                           &     0                            &     $\pi$                        \\
$m^*$ ($m_0$)                      &     0.066(5)                     &     0.033(2)                     \\
$v_F$ (10$^5$ m/s)                 &     4.44(7)                      &     7.66(4)                      \\
$\varepsilon_F$ (meV)$^\dag$       &     36.6(7)                      &     -110.5(8)                    \\
$n$ (10$^{18}$cm$^{-3}$)$^\ddag$   &     ($-$)10.8(2)                 &    ($+$)9.04(6)                  \\
$\mu$ (cm$^2$/Vs)$^\ddag$          &     1.9(2)$\times$$10^5$         &     1.9(3)$\times$$10^6$         \\
$T_D$ (K)                          &     3.4(4)                       &     30(2)                        \\
$\tau_{tr}$ (s)                    &     7.1(3)$\times$$10^{-12}$     &     3.7(2)$\times$$10^{-11}$     \\
$\tau_{Q}$ (s)                     &     3.6(4)$\times$$10^{-13}$     &     3.8(5)$\times$$10^{-14}$     \\
$l_{tr}$ ($\mu$m)                  &     3.2(7)                       &     27(3)                        \\
\end{tabular*}
{\rule{\temptablewidth}{1pt}}
\end{center}
\vspace*{-18pt}
\begin{flushleft}
$^\dag$ The Fermi energy for the $\alpha$-pocket is referenced from the bottom of the parabolic conduction band (electron-type), and for the $\beta$-pocket it is referenced from the Weyl point (hole-type).\\
$^\ddag$ Estimated from the Hall effect.\\
\end{flushleft}
\end{table}

The absence of spatial inversion symmetry in NbAs is reminiscent of the antisymmetric spin-orbit coupling that potentially results in a Rashba semiconductor, in which the FS consists of two pockets, an inner FS (IFS) and an outer FS (OFS). However, this seems to be unlikely in NbAs. First, we observed two different Berry phases for $\alpha$- and $\beta$-pockets, demonstrating that the two pockets are of different topological origins. This is apparently in contrast with the Rashba semiconductors, e.g., BiTeI\cite{Murakawa-Rashba} and BiTeCl\cite{Xiang-BiTeCl}, in which both of the IFS and OFS are topologically nontrivial. Second, since the system possesses a mirror plane $M_x$ (and $M_y$)\cite{Weng-TmPn}, the Rahsba picture expects a spin-degenerate FS in the configuration of $\textbf{B}\parallel\textbf{a}$ ({\it i.e.}, $\theta=$ 90\textordmasculine); whereas, we detect that the two FFT peaks split into four peaks in this configuration (Fig.~\ref{Fig.5}). Since the $\beta$-pocket is both smaller than the $\alpha$-pocket and has a nontrivial Berry phase, it is reasonable to assign the $\beta$-pocket as minority hole-like with higher mobility and the $\alpha$-pocket as the majority electron-like.



In conventional transport theory, a large magnetoresistance stems from a multi-band effect: although no net current flows in the $\textbf{y}$-direction, the currents carried in the $\textbf{y}$-direction by a particular type of carrier may be non-zero\cite{Singleton-Band}. These transverse currents experience a Lorentz force that is antiparallel to the $\textbf{x}$-direction. This backflow of carriers provides an important source of magnetoresistance which is most pronounced in semimetals like Bi\cite{Alers-BiMR} and WTe$_2$\cite{Ali-WTe2XMR,LuoY-WTe2Hall} where electrons and holes are compensated. This electron-hole compensation effect is also possible in NbAs, considering the close values of $n_e$ and $n_h$ aforementioned. Recently, giant magnetoresistance was also observed in several topological materials, e.g., the 3D Dirac semimetal Cd$_3$As$_2$\cite{Ong-Cd3As2} and the 3D Weyl semimetal TaAs\cite{Zhang-TaAsSdH}. In these topological semimetals, suppression of backscattering results in a transport lifetime much longer than the quantum lifetime. The lifting of this protection by an applied magnetic field leads to a very large magnetoresistance\cite{Ong-Cd3As2}. According to the aforementioned experimental results, we may infer the transport lifetime for the two observed pockets, $\tau_{tr}^{\alpha}=\frac{m^*_{\alpha}\mu_{\alpha}}{e}=7.1\times10^{-12}$ s, and $\tau_{tr}^{\beta}=3.7\times10^{-11}$ s. The ratio $\tau_{tr}^{\beta}/\tau_{Q}^{\beta}\sim 1000$, characteristic of a severe reduction in backward scattering. In comparison, for the $\alpha$-pocket, this ratio is only $\sim$ 20, demonstrating little topological protection. Therefore, NbAs seems to host a combination of the two mechanisms: a novel compensation effect between topologically trivial electrons and nontrivial hole; in particular, the nontrivial holes are topologically protected. 

Finally, we place NbAs in the context of the other $TmPn$ Weyl semimetals. The global electronic structure and presence of Fermi arcs states are common to all four compounds\cite{Weng-TmPn,Lv-TaAsPRX,XuS-TaAsARPES,XuS-NbAsARPES}. However, clear variations exist in the low energy electronic structure, which is dictated on the precise location of the Weyl nodes and their distance from the chemical potential. For instance, in TaAs one observes small extremal orbits\cite{Huang-TaAsLMR}, but in TaP the SdH frequencies are much larger\cite{HuJ-TaPSdH} which hence require larger fields to reach the quantum limit. Recently, systematic first-principles calculations on the $TmPn$ family have been performed by Lee {\it et al}\cite{LeeC-TmPnband}. The electronic structure consists of four pairs of Weyl nodes for $k_z$=0 labeled W1 and eight pairs of Weyl nodes off the $k_z$=0 plane labeled W2. They found that in TaAs both types of Weyl nodes lead to topologically non-trivial orbits, but in all other members only trivial orbits exist (Of course, there are also additional trivial orbits that are not related to the Weyl nodes in all these four materials). While this contradicts the observations of a non-trivial Berry phase in our NbAs and a chiral anomaly induced negative longitudinal magnetoresistance found in the other family members {\it e.g.} TaP\cite{Shekhar-TaPLMR}, it illustrates the potential variation between compounds in this family.

We have performed first principles calculations of NbAs using the WIEN2K code\cite{Wien2k} with the exchange correlation potential of Perdew-Burke-Ernzerhof\cite{PBE}, spin orbit coupling, and the reported crystalline structure\cite{Furuseth-NbAs}. Our band structure is in good agreement with that reported in Ref.~\cite{LeeC-TmPnband}, with two spin-orbit split hole bands and electron bands crossing the chemical potential. We find extremal orbits of the two hole bands to be 5.2 and 25.3 T, while the extremal orbits for the electron bands are 32.7, 17.9 and 6.0 T. There are three frequencies for the electron orbits because one band crosses the chemical potential in two distinct positions in the Brillouin zone. While these (highest) frequencies are close to the observed frequencies, we note that all orbits here are due to topologically trivial Fermi surfaces that either encloses a pair of Weyl nodes or none at all. Furthermore, simply shifting the chemical potential so that the chemical potential forms non-trivial orbits could not be reconciled with the Fermi energy of more than 100 meV found for our $\beta$-pocket. Clearly, calculations that go beyond the local density or generalized gradient approximation perhaps including a small degree of electronic correlations are required to find satisfactory quantitative agreement between theory and experiment at the level of tens of meV.

\section{Conclusion}

In conclusion, we have measured the electronic structure of NbAs via Shubnikov-de Haas oscillations in the transverse magnetoresistance. We find evidence for two pockets, which on the basis of our Hall effect measurements have opposite carrier types. We assign the larger $\alpha$-pocket as electron-like, and the smaller $\beta$-pocket we assign as hole-like. From a Landau fan analysis we find that the $\alpha$-pocket is topologically trivial, but the $\beta$-orbit is not. The Weyl node in NbAs is estimated from our analysis to lie 110.5 meV above the chemical potential. Finally, the transport lifetime of the holes on the $\beta$-pocket is three orders in magnitude longer than the quantum lifetime emphasizing the potential for generating technologically useful devices based on topologically non-trivial materials. We provide a novel case of electron-hole compensation where the giant magnetoresistance is caused by a cooperation of electron-hole compensation and a topological protection mechanism.

\section*{Acknowledgments}

We thank Philip Moll, James Analytis, Brad Ramshaw, and Yaomin Dai for insightful conversations, and V. S. Zapf for technical support. Samples were synthesized and characterized under the auspices of the Department of Energy, Office of Basic Energy Sciences, Division of Materials Science and Engineering. Electrical transport measurements and electronic structure calculations were supported by the LANL LDRD program. Work at the NHMFL  Pulsed  Field  Facility  is  supported  by  the  National  Science  Foundation,  the  Department  of  Energy, and the State of Florida through NSF cooperative grant DMR-1157490. Y. Luo acknowledges a Director's Postdoctoral Fellowship supported through the LANL LDRD program.


%

\end{document}